\documentclass[aps,letterpaper,twocolumn,preprintnumbers,floatfix,superscriptaddress]{revtex4-1} 

\usepackage{amsmath}
\usepackage{amsfonts}
\usepackage{amssymb}
\usepackage{graphicx, rotating}
\usepackage{epstopdf}
\usepackage{epsfig}
\usepackage{latexsym}
\usepackage{color}
\usepackage[dvipsnames]{}
\usepackage{multirow}
\usepackage{xr}
\usepackage{booktabs}
\usepackage[table]{}
\usepackage{xcolor,colortbl}
\usepackage{float}

\usepackage{slashed}
\usepackage{hyperref}
\hypersetup{colorlinks, citecolor=bluscuro, linkcolor=bluscuro, urlcolor=bluscuro}
\definecolor{rossos}{cmyk}{0,1,1,0.55}
\definecolor{bluscuro}{rgb}{0.15, 0.2, .85}
\definecolor{bluchiaro}{cmyk}{1,.3,0.,0.1}
\definecolor{verdescuro}{rgb}{0.3,0.8,0.3}

\newcommand{\be}{\begin{equation}}
\newcommand{\ee}{\end{equation}}          
\newcommand{\bea}{\begin{eqnarray}}
\newcommand{\eea}{\end{eqnarray}}

\begin{document}

\widetext

\title{
Gravitational Wave Scattering via the Born Series:  
\\ Scalar Tidal Matching to ${\cal O}(G^7)$ and Beyond}

\author{Simon Caron-Huot}
\author{Miguel Correia}
\affiliation{Department of Physics, McGill University, 3600 Rue University, Montr\'eal, H3A 2T8, QC Canada}
\author{Giulia Isabella}
\author{Mikhail Solon}
\affiliation{Mani L. Bhaumik Institute for Theoretical Physics, Department of Physics and Astronomy, University of California Los Angeles, Los Angeles, CA 90095, USA}

\begin{abstract}
\noindent 
We introduce a novel method to compute gravitational wave amplitudes within the framework of effective field theory. By reinterpreting the Feynman diagram expansion as a Born series, our method offers several key advantages. It directly yields  partial wave amplitudes, streamlining the matching with black hole perturbation theory. Long-distance gravitational interactions are unambiguously factorized from short-distance tidal effects, including dissipation, which are systematically incorporated via an in-in worldline effective action. 
Crucially, at every order in perturbation theory, integrals are expressed in terms of harmonic polylogarithms, enabling an end-to-end computation scalable to arbitrary orders. We illustrate the method with new predictions for scalar black hole Love numbers and their Renormalization Group equations to \(\mathcal{O}(G^7)\).

 \end{abstract}

\maketitle
\medskip

\noindent \textit{Introduction.}  The advent of gravitational wave astronomy has led to the observation of black hole and neutron star binaries \cite{LIGOScientific:2016aoc,LIGOScientific:2017vwq}, driving a surge in the application of field-theoretic methods to describe these systems \cite{Buonanno:1998gg,Buonanno:2000ef,Goldberger:2004jt,Goldberger:2005cd,Goldberger:2009qd,Porto:2016pyg,Rothstein:2014sra, Kalin:2020mvi,Mogull:2020sak,Cheung:2018wkq,Kosower:2018adc,Buonanno:2022pgc,Goldberger:2022ebt}. The finite-size nature of such compact objects is characterized by their Love numbers, which quantify the tidal response to an external gravitational field \cite{Binnington:2009bb,Damour:2009vw}. Crucially, Love numbers leave a distinctive imprint on the gravitational wave signal \cite{Flanagan:2007ix,Hinderer:2007mb,Hinderer:2016eia} and have already been constrained for neutron stars \cite{De:2018uhw,LIGOScientific:2018cki,LIGOScientific:2018hze}. Within the framework of effective field theory (EFT), Love numbers are rigorously defined as the Wilson coefficients of operators in the worldline effective action of the compact object \cite{Goldberger:2004jt,Goldberger:2005cd,Goldberger:2009qd,Porto:2016pyg}. Interestingly, the static Love numbers of black holes vanish identically \cite{Binnington:2009bb,Damour:2009vw,Kol:2011vg,Porto:2016zng,Charalambous:2021kcz, Hui:2021vcv,Charalambous:2022rre,Ivanov:2022qqt,Goldberger:2022ebt}.

In recent work \cite{Ivanov:2022qqt, Ivanov:2022hlo, Saketh:2023bul, Ivanov:2024sds, Saketh:2024juq}, Love numbers were determined by comparing scattering amplitudes in the EFT to black hole perturbation theory (BHPT) \cite{1973ApJ...185..635T,1973ApJ...185..649P,1974ApJ...193..443T,Mano:1996vt,Mano:1996gn,Mano:1996mf,Sasaki:2003xr,Bonelli:2022ten,Dodelson:2022yvn,Aminov:2023jve,Aminov:2024mul}. Crucially, scattering amplitudes are gauge invariant, and linking them to Love numbers provides a robust framework that avoids ambiguities inherent in coordinate-dependent approaches. Nonetheless, it remains a challenge to systematically compute scattering amplitudes to high orders in perturbation theory; for instance, matching the leading dynamical Love number in gravity requires a five-loop calculation at $\mathcal{O}(G^6)$ in Newton's constant  $G$, which is far beyond the current state-of-the-art at two-loops \cite{Ivanov:2024sds}.

In this Letter we introduce a new method that enables the EFT computation of gravitational wave amplitudes and consequent matching of tidal Love numbers at the required high orders in perturbation theory. We study scalar wave scattering and present new predictions for scalar black hole Love numbers at ${\cal O}(G^7)$, reproducing the $O(G^3)$ results of \cite{Ivanov:2024sds}.

To study the tidal responses of a compact object of mass $M$, we consider the scattering of a wave with frequency $\omega$ in the regime
\be
\label{eq:Bornregime}
G M \omega \, \ll \,1 \, \ll \,GM^2 \, , \qquad (\hbar = c = 1) \,.
\ee
The left inequality means that the probe's wavelength is much larger than the Schwarzschild radius, while the right inequality is the classical limit, i.e., the limit of large gravitational charge $M \gg M_{\rm Planck}$. As noted in \cite{Correia:2024jgr}, the Feynman diagram expansion in this regime takes the form of a Born series
\begin{figure}[H]
    \centering
\includegraphics[scale=0.799]{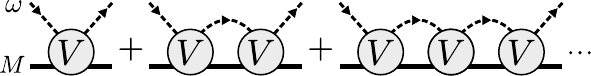}
    \label{Bornseries}
\end{figure}
\vspace{-5mm} \noindent
where the \emph{potential} $V$ collects two-particle irreducible diagrams in a perturbative series in $G$. In our case, $V$ includes the expected gravitational long-distance interactions but crucially differs from it in the region near the compact object, which is replaced by effective contact interactions that model tidal effects. These are precisely the object's Love numbers.  

\vspace{8pt}

\noindent \textit{Setup -- The Effective Wave Equation.} Due to the connection with the Born series, we can interpret the scattering amplitude calculated in the EFT as arising from an underlying wave equation, which we refer to as the \emph{effective wave equation}. This equation can be derived in the standard manner from an effective action, where the parameters are determined by matching the physical scattering phases to those of the complete underlying theory. In our scalar model, given that $\omega \ll M$
as indicated by \eqref{eq:Bornregime}, the scalar field $\phi$  does not produce its own gravitational field. As a result, its long-distance dynamics are governed solely by the action
\begin{equation}\label{eq:Sgrav}
   \, S_\text{Grav} = -{1 \over 2} \int \mathrm{d}^d x \sqrt{-g}\, g^{\mu \nu} \partial_\mu \phi \, \partial_\nu \phi \,,
\end{equation}
where $g^{\mu\nu}$ is the Schwarzschild metric in $d$-spacetime dimensions,
\begin{equation}
\label{eq:metric}
    ds^2 = - f dt^2 + f^{-1} dr^2 + r^2 d\Omega^2_{d-2}
\end{equation}
with
\begin{align}
  \!\! \!\!\! f(r) = 1 - {2 G M  \, n_d \,\mu^{4-d} \over r^{d-3}}\,, \quad\!\!\!\! n_d \equiv {4 \pi^{3 - d \over 2} \Gamma\big( {d -1 \over 2}\big) \over d - 2}\,.
\end{align} 
The metric $g^{\mu\nu}$ is understood to be perturbatively expanded in Newton's constant $G$, whose units are kept as in four-dimensions via the introduction of the dimensionful scale $\mu$. This allows us to accurately replicate the Feynman diagram computation in dimensional regularization where as usual $d = 4 - 2 \epsilon$.
As we will see, the perturbative diagrams are dominated by the single scale $r\sim \omega^{-1}\gg GM$ and sensitivity to the near-horizon region will appear in the form of ultraviolet (UV) divergences regulated by $\epsilon$.

We supplement \eqref{eq:Sgrav} with a worldline effective action modeling the tidal response of the compact object, 
\begin{align}\label{eq:SLove}
  \!\!\!  S_\text{Love}&=\sum_{\ell,n} \frac{\mu^{4-d}}{\ell!}\int \mathrm{d}\tau \Big[C_{\ell,n}(\nabla_{\!\ell} \,\phi_+)\partial_\tau^n(\nabla_{\!\ell} \,\phi_-)  \Big]
\end{align}
where the spatial derivative is defined as $\nabla^\mu = (g^{\mu\nu}+u^\mu u^\nu) \nabla_\nu$, and $u^\mu=(1,0,0,0)$ is the 4-velocity of the compact object. The symbol $\nabla_\ell$ denotes the symmetrized and trace-subtracted product of $\ell$ spatial derivatives. The coefficients $C_{\ell,n}$ are the Love numbers,
and we again inserted a power of the scale $\mu$ to ensure that their units are as in four-dimensions: $[C_{\ell,n}]=-2\ell-n-1$.
Being supported on $\mathbf{r}=0$, where $x^\mu = (t,\mathbf{r})$, the worldline action \eqref{eq:SLove} adds a potential proportional to $\delta^{d-1}(\mathbf{r})$  to the wave equation.

An essential aspect of tidal physics is dissipation.
As for other open systems, an action can be written using the in-in field doubling. We use the Keldysh basis \cite{Keldysh:1964ud,Martin:1973zz} in which $\phi_+\!=\!\frac12(\phi_1{+}\phi_2)$
represents the classical field and $\phi_-\!=\!\phi_1{-}\phi_2$ the source.
For conservative effects, the in-in action is simply the difference $S[\phi_1]{-}S[\phi_2]$ of two conventional actions (as in \eqref{eq:Sgrav}) 
\begin{align}\label{eq:SK}
\!\! S[\phi]\!\supset\!\!\!\int\!\! dt \frac{\phi \partial_t^n \phi}{2} \!\Rightarrow
 S_\text{in-in}\!\supset\!\frac{1{+}({-}1)^n}{2}\!\!\!\int \!\!dt\, \phi_+ \partial_t^n \phi_-
\end{align}
while the non-conservative Love numbers, $C_{\ell,n}$ with $n$ odd, can only be written in the form \eqref{eq:SLove}.
We will focus on the retarded two-point amplitude $\langle \phi_- \phi_+\rangle$ which satisfies the same boundary conditions as the classical problem and reduces to it in the classical limit \cite{Caron-Huot:2023vxl,Biswas:2024ept} (see Appendix \ref{app:iepsilon}).

The effective wave equation is the equation of motion derived from varying $\phi_-$ in the action $S= S_{\rm Grav,SK} + S_{\rm Love}$. After projecting the scalar field onto spherical harmonics $\phi_+(\mathbf{r}) = {\psi(r) \, Y^m_\ell(\hat{\mathbf{r}}) \over r^{(d-2)/2}}$ we find
\begin{equation}
\label{eq:wavedimreg}
    \!\!\!\!\!\left[ {d^2 \over dr^2} - {(\ell - \epsilon) (\ell - \epsilon +1) \over r^2} + \omega^2 \right]\! \psi(r) = V(r) \psi(r)
\end{equation}
where the potential is given by $V(r)=V_{\mathrm{Grav}}(r)+ V_{\mathrm{Love}}(r)$, and $\ell$ is the total angular momentum eigenvalue.
The first term is the long-distance gravitational potential derived from $S_{\rm Grav}$ in  $(\ref{eq:Sgrav}$), 
\begin{align}\nonumber
   & V_{\mathrm{Grav}}(r) =   \sum_{n = 1}^\infty  \left( {2 G M n_d \,\mu^{2\epsilon}\over r^{1- 2 \epsilon}}\right)^{\!n}  \Big[ {2 \epsilon -1 \over r} {d \over dr} \\&
    +{\ell^2 + \ell + 1 - \epsilon(3 + 2 \ell) + 2 \epsilon^2 \over r^2} -(n+1) \, \omega ^2 \Big]\label{eq:Vgrav} \,,
\end{align}
which is known to  reproduce the family of ``fan" Feynman diagrams \cite{Duff:1973zz,Damgaard:2024fqj,Kosmopoulos:2023bwc, Cheung:2023lnj} 
\begin{figure}[H]
    \centering
\includegraphics[scale=1.015]{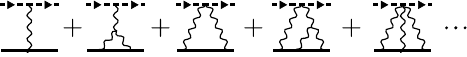}
    \label{Vgrav}
\end{figure}
\vspace{-5mm} \noindent The second term is the short-distance potential due to the wordline effective action $S_{\rm Love}$ in (\ref{eq:SLove}), 
\begin{equation}\label{eq:VLove}
    V_{\mathrm{Love}}(r) =
    -\mu^{2\epsilon}\sum_\ell F_{\ell}(\omega) \frac{(-1)^{\ell}}{\ell!}\nabla_{\!\!\ell}\,\delta^{d-1}(\mathbf{r})\,\nabla_{\!\!\ell}\,,
\end{equation}
with
\begin{equation}
\label{eq:Fl}
    F_\ell(\omega)=
    \sum_{n=0}^\infty \, (i \omega)^n\,C_{\ell,n},  \;\;\;\; C_{\ell,n}= \raisebox{-1.4em}{\includegraphics[width=1.2cm]{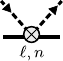}}\,.
\end{equation}

\vspace{8pt}

\noindent \textit{Born Series, $S$-matrix and Boundary Conditions.} The radial wave equation \eqref{eq:wavedimreg} can be solved perturbatively via the Born series as in quantum mechanics,
\begin{align}\label{eq:psiBorn}
    &\psi(r) = \Psi_0(r) + \int\displaylimits_0^\infty \mathrm{d}r_1\mathrm{G}(r,r_1)V(r_1)\Psi_0(r_1)\\\nonumber
    + &\int\displaylimits_0^\infty \mathrm{d}r_1\mathrm{G}(r,r_1)V(r_1)\int\displaylimits_0^\infty \! 
\mathrm{d}r_2\,\mathrm{G}(r_1,r_2)V(r_2) \Psi_0(r_2) \dots
\end{align}
where the Green's function \( \mathrm{G}(r, r') \) and the free solution \( \Psi_0(r) \) satisfy the free radial wave equation given by \eqref{eq:wavedimreg} when $V(r) = 0$. The free solution \( \Psi_0(r) \) is a linear combination of Riccati-Hankel functions \cite{Taylor:1972pty},
 \begin{equation}
\Psi_0(r) = A_\text{in} \, h_{\ell - \epsilon}^-( \omega r) + A_\text{out} \, h_{\ell - \epsilon}^+( \omega r) \,,
 \end{equation}
 which behave as free radial waves at infinity,
 \begin{equation}\label{eq:metric}
     h_{\ell - \epsilon}^\pm (r \to \infty)
    = e^{\pm i\left( \omega r-\frac{\pi}{2}(\ell-\epsilon)\right)}\,
 \end{equation}
As is customary in scattering theory \cite{Taylor:1972pty,Weinberg:1995mt}, we assume that interactions decay sufficiently rapidly at infinity, \( V(r \to \infty) \lesssim 1/r \), ensuring that \( \psi(r \to \infty) \approx \Psi_0(r) \). The \( S \)-matrix, or more precisely, the partial wave amplitude, is defined as the ratio of the coefficients of the outgoing and ingoing free waves,
\begin{equation}\label{eq:Smatrix}
     S_\ell(\omega) \equiv -\frac{A_\mathrm{out}}{A_\mathrm{in}} \,.
 \end{equation}
Moreover, it is also assumed that the centrifugal potential ${(\ell-\epsilon)(\ell-\epsilon+1) \over r^2}$ dominates at small distances  over the interaction potential $V(r \to 0)$, so that $\psi(r \to 0) \approx \Psi_0(r)$. The Riccati-Hankel functions in this limit behave as linear combinations of ``regular" and ``irregular" terms
\begin{equation}\begin{aligned}
\label{eq:hp}
 h_{\ell - \epsilon}^+ (r \to 0) \;-\;  h_{\ell - \epsilon}^- (r \to 0) \;&\propto  \;r^{1+ \ell-\epsilon}\,,\\
 h_{\ell - \epsilon}^+ (r \to 0)\; +\;  h_{\ell - \epsilon}^- (r \to 0)\; &\propto \;r^{\epsilon-\ell} \, .
\end{aligned}\end{equation}
At $r \to 0$ the solution then assumes the form
\be
\psi(r \to 0) = B_\mathrm{reg} \, r^{1+ \ell-\epsilon} + \frac{B_\mathrm{irr} \, \mu^{2\epsilon}}{2\ell+1-2\epsilon} r^{\epsilon-\ell}\,. \label{eq:psi0near}
\ee
When $V\!=\!0$ the $\phi_+$ equation of motion at the origin
imposes the boundary condition $B_\text{irr} = 0$.
This can be seen from the fact that the irregular solution produces a $\delta$-function term (recall that $\phi_+\propto \Psi/r^{\frac{d-2}{2}}$):
\be
 \vec{\nabla}^2 \frac{r^{\epsilon}}{r^{\frac{d-2}{2}}}
 = - \frac{4\pi^{\frac{d-1}{2}}}{\Gamma(\frac{d-3}{2})}
\delta^{d-1}(\mathbf{r})\,.
\ee
Using this identity and its derivatives, it is easy to see
that canceling the $\delta$-function in the equation of motion including the Love number potential $V_\text{Love}$ now imposes a modified boundary condition relating the two coefficients, where using  \eqref{eq:VLove}
we find specifically:
\begin{align}
\label{eq:BLove}
\frac{B_\text{irr}}{B_\text{reg}}=c_\ell\, F_\ell(\omega),\,\,\,\,
 c_{\ell}=
\frac{2^\ell}{2\pi ^{\frac{3}{2}-\epsilon}}\Gamma\Big(\frac{3}{2}+\ell-\epsilon\Big).
\end{align}
This closely resembles the definition of Love numbers found in the GR literature \cite{Damour:2009vw,Binnington:2009bb, Cardoso:2017cfl}.
The advantage here is that $F_\ell(\omega)$ has a gauge-invariant definition based on the worldline action \eqref{eq:SLove}. Equation \eqref{eq:BLove} thus connects the GR and EFT definitions of Love numbers, which is one of our main conceptual results.  Because it allows to trade $V_\text{Love}(r)$ for a re-definition of the boundary condition at $r \to 0$, it unambigously factorizes short-distance and long-distance effects.

It may seem surprising that the gravitational potential $V_\text{grav}$ does not modify this boundary condition.
An intuitive explanation is that in dimensional regularization, 
the $r\to 0$ behavior is in effect regulated by treating $r^{-1+2\epsilon}$ as small when $r\to 0$.

In the free case (\( V(r) = 0 \)), the boundary condition $B_\text{irr}=0$ imposes that $A_\text{in}=-A_\text{out}$ via the relation \eqref{eq:hp}. Consequently, this implies \( S_\ell(\omega) = 1 \), as expected. In the general case, however, the relationship between \( A_\text{out} \) and \( A_\text{in} \) is more intricate and can be determined perturbatively in the potential \( V(r) \) using the Born series \eqref{eq:psiBorn}.

The assumptions that the interacting solution \( \psi(r) \) reduces to the free solution \( \Psi_0(r) \) both at infinity (\( r \to \infty \)) and at the origin (\( r \to 0 \)), while satisfied within dimensional regularization, are in fact \emph{not} satisfied when $\epsilon = 0$.
This is the fundamental reason why the Born series and the partial wave amplitude $S_\ell(\omega)$ exhibit IR and UV divergences which we need to regularize by keeping $\epsilon \neq 0$, in close parallel with the Feynman diagram computation \cite{Ivanov:2024sds}. The IR divergences are entirely due to the Coulomb term $n=1$ in $V_\text{Grav}(r)$ and obey a very simple exponentiation \cite{Taylor:1972pty}.
The UV divergences, on the other hand, which must cancel
between $V_{\rm Love}$ and $V_{\rm Grav}$ contributions, originate from structure at the Schwarzschild radius which is not resolved by the perturbative expansion in the region $r\sim \omega^{-1}\gg GM$.

Concretely, the relationship between partial waves \( S_\ell(\omega) \) and momentum-space Feynman diagrams involves an integral over Legendre polynomials 
\cite{Ivanov:2024sds}. Notably, the EFT matching with BHPT is performed at the level of the partial waves \cite{Ivanov:2024sds} and, crucially, our method yields directly the partial wave \( S_\ell(\omega) \) within the EFT framework, eliminating the need for this intermediate step.

\vspace{8pt}
\noindent \emph{General Strategy and Near-Far Factorization.} With the boundary conditions and the connection to the amplitude in \eqref{eq:Smatrix} established, we turn to the computation of the iterated integrals in the Born series \eqref{eq:psiBorn}, where $V(r) = V_\text{Grav}(r) + V_\text{Love}(r)$ are given by \eqref{eq:Vgrav} and \eqref{eq:VLove}. As mentioned, some integrals diverge when $\epsilon \to 0$. Unfortunately, the Riccati-Hankel functions $h_{\ell - \epsilon}^\pm(\omega r)$, from which the Green's function $\mathrm{G}(r,r')$ can be built of, do not have a closed form in terms of elementary functions for $\epsilon \neq 0$, making these integrals hard to solve in full generality.

Our strategy is to separate the problem in two different regions which we refer to as \emph{far-zone} and \emph{near-zone} where
we can make distinct approximations.
The idea is similar to the method of ``matched asymptotic expansions'' \cite{bender1999advanced} and the ``method of regions" in the context of Feynman integrals \cite{Smirnov:2002pj}. 

In the far-zone region we choose the Green's function $\mathrm{G}^\text{far}(r,r')$ to only have support for ordered radii $0{<}r{<}r'$. This eliminates any possible UV divergence from $r'=0$ and allows for a regular expansion in $\epsilon$ where the starting point are Riccati-Hankel functions $h_{\ell}^\pm(\omega r)$ that reduce  to  elementary functions for integer $\ell$. Moreover, in the far-zone region, the Love numbers drop out out of the calculation due to their compact support at $\!r'=\!0$. The iterated integrals over $V_\text{Grav}(r)$ which contribute to $S_\ell(\omega)$ can then be conveniently expressed in terms of harmonic polylogarithms (HPLs), as we detail in the next section. 

In the near-zone region, on the other hand, we choose a Green's function $\mathrm{G}^\text{near}(r,r')$ that only has support for $r'{<}r$. Here we have to keep $\epsilon$ finite in order to regulate UV divergences and simplifications come from the fact that the Riccati-Hankel functions $h_{\ell-\epsilon}^\pm(\omega r)$ at small $r$ (or $\omega$) admit simple power series starting 
with the ``regular" and ``irregular" terms in \eqref{eq:hp}.
The near-zone Born series then proceeds by starting with the boundary condition \eqref{eq:psi0near} incorporating the Love numbers via \eqref{eq:BLove}, and iteratively integrating over $V_{\rm grav}$ and $\mathrm{G}^\text{near}$.

When the dust settles we have two solutions of the same differential equation in two different regions. They are made to match by bringing the far-zone solution to $r \to 0$ and expanding the near-zone solution in $\epsilon$. This leads to the linear relation
\begin{gather}\label{eq:fact}
 \begin{bmatrix}  A_{\mathrm{in}} \\  A_{\mathrm{out}} \end{bmatrix}
 \propto
 \mathrm{W}_\text{far-near} \cdot
   \begin{bmatrix}  1 \\ c_\ell F_\ell(\omega)  \end{bmatrix} 
\end{gather}
where $\mathrm{W}_\text{far-near}$ is the Wronskian matrix \cite{bender1999advanced} that connects the two independent far-zone and near-zone solutions at $r \to 0$ and whose calculation depends only on $V_\text{Grav}(r)$. Equation \eqref{eq:Smatrix} then determines the partial wave amplitude $S_\ell(\omega)$ in terms of the Love numbers in $F_\ell(\omega)$, enabling the latter to be fixed  by matching $S_\ell(\omega)$ to the BHPT answer \cite{1973ApJ...185..635T,1973ApJ...185..649P,1974ApJ...193..443T,Mano:1996vt,Mano:1996gn,Mano:1996mf,Sasaki:2003xr,Bonelli:2022ten,Dodelson:2022yvn,Aminov:2023jve}.

{\def\arraystretch{2}\tabcolsep=13pt
\newcolumntype{?}{!{\vrule width 0.7pt}}
\begin{table*}[hbt!]
  \centering
  \begin{tabular}{c?c|c|c} 
  & $\ell=0$ & $\ell =1$ &  $\ell = 2$  \\ \midrule[0.7pt]
     $G$ & 0 &-- &--\\ \hline
    $G^2$ & $4\pi R_s^2$ & -- & --\\\hline
    $G^3$ & $4\pi R_s^3\left( \frac{1}{4\epsilon}+\log{R_s\bar\mu}+\frac{19}{12}+\gamma_E\right)$ & 0 & --\\\hline
     $G^4$ & $\frac{44}{3}\pi R_s^4\left( \frac{1}{4\epsilon}+\log{R_s\bar\mu}-\frac{13}{66}-\frac{\pi^2}{11}+\gamma_E\right)$ & $\pi \frac{R_s^4}{3}$ &-- \\\hline
      $G^5$ & $\!\!\!\!\frac{11}{6}\pi R_s^5\,\Big[\frac{1}{4\epsilon^2}+\frac{2}{\epsilon}\left(\log{R_s\bar\mu}+\gamma_E+\frac{56}{33} \right)+ $ & $\frac{1}{6}\pi\, R_s^5\left( \frac{1}{4\epsilon}+2\log{R_s\bar\mu}-\frac{19}{30}+2\gamma_E\right)$ & 0 \\[-0.2cm]
      &$\!\!\!\!\!\!\left(2\log{R_s\bar\mu}+2\gamma_E+\frac{167}{66} \right)^2-\frac{338611}{21780} - \frac{8 \pi^2}{3} - \frac{48 \,\zeta_3}{11}\Big]$ & & \\
  \end{tabular}
  \caption{Scalar black hole Love numbers $C_{\ell,n}$ in dimensional regularization where $R_s=2GM$ and $\bar\mu^2 = \mu^2 4\pi e^{-\gamma_E}$. 
  }
  \label{tab:Love}
\end{table*}}

\vspace{8pt}

\noindent \textit{Far-zone Born Series and Harmonic Polylogarithms.} In the far-zone we perturbatively expand the Born series \eqref{eq:psiBorn} in a double series in $G$ and $\epsilon$, treating  ${\epsilon( 1 + 2 \ell - \epsilon) \over r^2} + V_\text{Grav}(r)$ as the perturbing potential
where $V_\text{Grav}(r)$ given in \eqref{eq:Vgrav} is expanded in $\epsilon$. The solution $\Psi_0(r)$ for $\epsilon=0$ is a linear combination of Riccati-Hankel functions $h^\pm_\ell(\omega r)$ with integer argument $\ell$,
\begin{equation}
\label{eq:hpm}
  \!\!  h_\ell^{\pm}(z) = e^{\pm i z} \sum_{n = 0}^\ell {a^\pm_{n,\ell}  \over z^n}, \quad \!\!\! a^\pm_{n,\ell} =  {(\pm i)^{n-\ell} (\ell + n)! \over  2^n \; n! (\ell - n)!}
\end{equation}
and the Green's function is given by (with $z=\omega r$)
\begin{equation}
\!\!\!\mathrm{G}^{\rm far}
(r,r')\!=\!{h_\ell^+(z) h_\ell^-(z') 
-(z{\leftrightarrow}z') \over 2 i \omega} \Theta(r'-r).
\end{equation}
\noindent 
It is crucial for our method that \eqref{eq:hpm} consists of simple exponentials times polynomials in $z^{-1}$.
With the use of integration-by-parts identities the far-zone Born integrals \eqref{eq:psiBorn} are expressible in terms of iterated integrals defined recursively by

\begin{equation}\label{eq:Irecursion}
{\cal I}_{\ldots, j_n}(z_{n+1}) \!\equiv\!-\!\!\int_{z_{n+1}}^\infty \!\!\frac{dz_n}{z_n} e^{\sigma_n z_n} \frac{\log(\frac{z_{n+1}}{z_{n}})^{j_n{-}1}}{(j_n{-}1)!}
{\cal I}_{\ldots}(z_n)
\end{equation}
with seed ${\cal I}(z_1)=1$ and $\sigma_n=2i (-1)^{n-1}$. 
Because of the exponential factor, these integrals are \emph{not}
simply related to harmonic polylogarithms \cite{Remiddi:1999ew}.
Nonetheless, to compute the partial wave $S_\ell(\omega)$ via \eqref{eq:fact} we only need their limit as $r \to 0$ which remarkably \emph{can} be calculated using harmonic polylogarithms \footnote{A seemingly equivalent family of iterated integrals has been introduced recently in \cite{Aminov:2024aan}, who also related the complete integrals to multiple zeta values using a different method.}. To see this, we make the recursive change of variables
\be
\frac{z_{k+1}}{z_k} = {1 - u_{k} \over 1 - u_{k-1}} \left({u_{k-1} \over u_{k}} \right)^{\delta_{k,\text{even}}} ,
\ee
starting at $k = 1$ and setting $u_0 \equiv 0$, and where the Kronecker delta is nonzero when $k$ is even.
This change is designed to collects all exponentials into the outermost integral over $z_1$, which can then be readily performed as $z_{n+1}\to 0$ where it becomes a complete integral.
This limit introduces divergences which we regulate by replacing
the innermost logarithmic
power in \eqref{eq:Irecursion} by $\frac{(z_{n})^\eta}{\eta^{j_n-1}}$.
The complete regulated integral is then expressed as an integral over
the region $\Gamma = \{0<u_1<\dots<u_{n-1}<1\}$,
\begin{align}\nonumber
\!\!{\cal I}_{j_1\ldots j_n}^{(\eta)} \!\!&=\!\mathcal{N}\!\!\int\displaylimits_\Gamma\mathrm{d}\! \log{\left(1{-}u_1\right)}\mathrm{d}\!\log{\left(\!\frac{1{-}u_2}{u_2}\!\right)}\mathrm{d}\!\log{\left(1{-}u_3\right)}\cdots\\&
\!\!\!\!\!\!\!\times \prod_{k=1}^{n-1}\frac{\log\!\left(\!\frac{z_{k+1}}{z_{k}}\!\right)}{(j_{k}{-}1)!}^{\! j_{k}-1}\!\!\!\frac{1}{(-\eta)^{j_n}}\left[
\frac{(u_{n-1})^{\delta_{n,\mathrm{even}}}}{1-u_{n-1}}\right]^\eta
\label{eq:Ieta} \end{align}
where $\mathcal{N}=\Gamma(1{+}\eta)(-2i)^{-\eta}$.
The $z$ integrals produce regularized HPLs that evaluate to multi-zeta values \cite{Remiddi:1999ew}, and the auxiliary regulator $\eta$ drops out once we return to the incomplete integrals ${\cal I}_{\ldots}(z)$ using the simple identity $\int_{r}^\infty=\int_0^\infty-\int_0^r$ (the far-zone solution is UV finite). Examples are given in Appendix \ref{app:intexample}.

Note that all integrals \eqref{eq:Irecursion} are IR finite. IR divergences in the far-zone Born series \eqref{eq:psiBorn} occur when the innermost integrand is of the form ${\log^nr \over r}$ for some integer $n$. In this case we trivially integrate up to a hard-cut-off $r < R$ until all leftover integrals take the form \eqref{eq:Irecursion}. We then express $R$ in terms of the dimensional regulator $\epsilon$ (see Appendix \ref{app:results}).

\vspace{8pt} 

\noindent \textit{Near-zone Born Series and Renormalization Group.} In the near-zone, we expand in $G$ and in $\omega$ (i.e. a small $r$ expansion). The potential that is iterated  by the Born series \eqref{eq:psiBorn} in the near-zone is $\omega^2+ V_\text{Grav}(r)$. The Green's function is given by 
\begin{equation}
\!\!\!\mathrm{ G}^\text{near}(r,r') \!=\!{ r^{\epsilon-\ell} (r')^{1 + \ell - \epsilon}- (r\,{\leftrightarrow}\,r')   \over {1+2 \ell-2 \epsilon }}\Theta(r-r')
\end{equation} 
with $\Psi_0(r)$ given by \eqref{eq:psi0near}.
All integrands in the near-zone are simple monomials and can be integrated trivially. 
A simple power counting argument may then be used to predict when UV divergences can first appear. For instance, 
from the first Born iteration $\int_0^r dr\, r^{2 + 2 \ell - 2 \epsilon} V_{\mathrm{Grav}}(r) \sim \int_0^r dr \, r^{2+ 2 \ell -n+2 \epsilon(n-1)}$ we see that a $1/\epsilon$ pole appears when $n=3+2\ell$ and, indeed, the first UV divergence shows up at order $\mathcal{O}(G^3)$ \cite{Ivanov:2024sds}. This also suggests an alternative way of regulating UV divergences by shifting $\ell$ away from integer values, as noticed in \cite{Dodelson:2022yvn,Bautista:2023sdf}.

The UV divergences are of course tightly related to a Renormalization Group (RG) equation which predicts logarithmic running of Love numbers.
All UV divergences in our approach originate from the boundary condition \eqref{eq:psi0near} becoming degenerate when $\epsilon=0$. 
By a change of basis
$(B_{\rm reg},B_{\rm irr})^T=Z(\bar{B}_{\rm reg},\bar{B}_{\rm irr})^T$
we can thus ensure that the solution expressed in terms of the \emph{renormalized coefficients} $\bar{B}$ remains finite as $\epsilon\to 0$.
According to standard renormalization theory, the price to pay is that coefficients become scale-dependent and satisfy a renormalization group equation: $(\mu\frac{d}{d\mu}+\gamma^{(\ell)})(\bar{B}_{\rm reg},\bar{B}_{\rm irr})^T=0$.
The matrix $Z$ and anomalous dimension in a minimal subtraction scheme are recorded in Supplementary Material \ref{app:results}. The renormalized Love numbers $c_\ell\bar{F}_\ell=\bar{B}_{\mathrm{irr}}/\bar{B}_{\mathrm{reg}}$ can then be interpreted as a boundary condition in the region $GM \,{\ll} \,r \,{\ll} \,\omega^{-1}$, providing a precise dictionary between the renormalization scheme used in the worldline EFT and the GR definition of Love numbers (see \eqref{psi0d4}). The renormalized Love numbers satisfy the RG equation: 
\begin{align}
\label{eq:RGE}
\!\!\!\!\mu \frac{d}{d\mu}\bar{F}_\ell(\omega)
=c_\ell\gamma^{(\ell)}_{1,2}\bar{F}_\ell^2+\big(\gamma^{(\ell)}_{1,1}-\gamma^{(\ell)}_{2,2}\big)\bar{F}_\ell-\frac{\gamma^{(\ell)}_{2,1}}{c_\ell}\,.
\end{align}
We note the presence of a quadratic term, which starts contributing at $\mathcal{O}(G^5)$ in the black hole case. To our knowledge it had not been identified before. 

\vspace{8pt}

\noindent \emph{Results.} In Table \ref{tab:Love} we show the Love numbers $C_{\ell,n}$ determined by matching our answer to BHPT, for $\ell=0,1,2$ up to $\mathcal{O}(G^5)$. Phase-shifts, RG equations, and Love numbers up to $\mathcal{O}(G^7)$ are listed in Appendix \ref{app:results}. We agree with the $\mathcal{O}(G^3)$ results of \cite{Ivanov:2024sds} and make new predictions through $\mathcal{O}(G^7)$.

\vspace{8pt}

\noindent \textit{Conclusion.} 
We anticipate that our method will enable the determination of the observationally relevant gravitational black hole Love numbers, which start at \(\mathcal{O}(G^5)\). 
Potential challenges include handling the graviton's spin-2 structure within dimensional regularization and accounting for gravitational backreaction on the worldline, which introduces additional contact terms. We also expect that the far-zone integrals, being expressible in terms of harmonic polylogarithms, can enhance BHPT computations \cite{Aminov:2023jve,Aminov:2024aan}. We plan to address these aspects in future work.

\vspace{8pt}

\noindent \emph{Acknowledgments.} We thank Zvi Bern, Vitor Cardoso, Matthew Dodelson, Liam Fitzpatrick, J.P. Gatica, Alessandro Georgoudis, Vasco Gon\c{c}alves, Tushar Gopalka, Enrico Hermann, Misha Ivanov, Callum Jones, Ami Katz, Julio Parra-Martinez, Michael Ruf, Filipe Serrano, Anna Wolz, Sasha Zhiboedov, and Zihan Zhou for useful discussions. 
The work of S.C.H. and M.C. is supported by the National Science and Engineering Council of Canada (NSERC) and the Canada Research Chair program, reference number CRC-2022-00421. M.S. and G.I. are supported by the US Department of Energy under award number DE-SC0024224, the Sloan Foundation and the Mani L. Bhaumik Institute for Theoretical Physics.

\appendix

\section{In-in and time-ordered amplitudes}
\label{app:iepsilon}

Dissipative effects on the compact object are naturally captured by the in-in (Schwinger-Keldysh) formalism. We highlight here similarities and distinctions with in-out or time-ordered amplitudes that have not been widely appreciated in the literature.
In the Keldysh basis the propagator coming from the free bulk action \eqref{eq:Sgrav} has two nonvanishing components, the retarded and symmetrized propagators
\begin{equation}\label{eq:props}\begin{split}
&\langle\phi_- \phi_+\rangle =\,\raisebox{-0.1em}{\includegraphics[width=1.7cm]{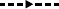}}\, = \frac{i}{(p_0+i\epsilon)^2-{\bf p}^2},\\
&\langle\phi_+ \phi_+\rangle =\, \raisebox{-0.4em}{\includegraphics[width=1.7cm]{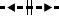}} \,= \pi \delta(p^2)\,.
\end{split}\end{equation}
The largest time equation implies that the
propagator between two $\phi_-$ fields vanishes
and that the full $\langle \phi_-\phi_+\rangle$ satisfies retarded boundary conditions. We use a graphical notation from \cite{Caron-Huot:2010fvq} which emphasizes the forward propagation of time toward $+$ vertices.

According to response theory \cite{Chou:1984es,Kapusta:2006pm}, the linear effect of an incoming wave on the expectation value of the far field is expressed by a commutator, here of asymptotic creation/annihilation operators \cite{Kosower:2018adc,Caron-Huot:2023vxl} in the presence of the black hole:
\begin{equation} \label{eq:BH exp val}
\!\! \!\!\!\!\langle{\rm BH}|\,[a^{\rm out}_{k'},a^{\rm in}_{k}]\,|{\rm BH}\rangle
\!=\\2\pi \delta(k'^0{-}k^0)2k^0
S^{\rm ret}(\mathbf{k}',\mathbf{k})
\end{equation}
where $a^{\rm out}=S^\dagger a^{\rm in} S$.
By a generalization of the LSZ reduction formula \cite{Caron-Huot:2023vxl}, the (connected part of this) \emph{retarded amplitude} is equal to the on-shell limit of the amputated retarded two-point function $\langle \phi_-\phi_+\rangle$.

It is easy to see that all diagrams contributing to \eqref{eq:BH exp val} consist of an uninterrupted sequence of retarded propagators and vertices, for example:
\begin{equation}
S^\text{ret} \;\supset \;\raisebox{-1em}{\includegraphics[width=2.4cm]{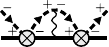}}\,\,\,.
\end{equation}
This is due to the vanishing of the $\langle \phi_-\phi_- \rangle$ propagator and the absence of a $\phi_+\phi_+$ term in the worldline action, both due to the largest time equation. This is a chief advantages of the Keldysh basis and ensures monotonic flow of time from source to observation.  Thus $S^\text{ret}$ is given by a standard Born series, albeit with retarded (causal) boundary conditions.
Fourier transforming the propagator to position space and projecting onto spherical harmonics, as described above \eqref{eq:wavedimreg}, yields the following radial Green's function \cite{Taylor:1972pty}
\begin{align}
\label{eq:pwGreen}
   & \langle\psi_-(r) \, \psi_+(r')\rangle ={ j_\ell(\omega r_<)\, h_\ell^+(\omega r_>) \over \omega}
\end{align}
where $r_< = \min(r,r'), \;\; r_> = \max(r,r')$ and $ j_\ell(z)=[h^+_\ell(z)-h_\ell^-(z)]/2i$ is the regular solution near the origin (see \eqref{eq:hp}). 
This confirms diagrammatically that in the absence of $V_\text{Love}$ the (dimensionally-regulated) Feynman diagram expansion constructs the purely regular solution near the origin, while the $V_{\rm Love}$ terms localize to $r'=0$ and source the irregular solution (see \eqref{eq:psi0near}).

The Green's function obtained from the Feynman rules, \eqref{eq:pwGreen},  differs from the ones used in the text by homogeneous solutions, for example:
\begin{align}
 \!\!\!\text{G}^{\rm far}(r,r')= \langle\psi_-(r) \psi_+ (r')\rangle \!-\!j_\ell(\omega r')h_\ell^+(\omega r)/\omega.
\end{align}
The choice made in the text simplifies the iterated integrals but ultimately leads to a multiple of the same solution once the boundary condition at the origin is accounted for.

For positive frequencies, the retarded Green's function \eqref{eq:props} coincides with the time-ordered one. In particular, the purely gravitational contribution to the retarded and time-ordered phase-shifts are identical ($\delta_{\rm Grav}$ below).  Could we not simply replace \eqref{eq:BH exp val} with the standard time-ordered product?

The answer is no, due to fluctuation effects on the worldline.
Dissipation in time-ordered perturbation theory could be captured by adding additional degrees of freedom (see e.g. \cite{Goldberger:2005cd}).
However, in analogy with e.g. Brownian motion \cite{Son:2009vu}, one should expect the time-ordered amplitude to probe additional fluctuations. The full worldline action can contain other vertices besides \eqref{eq:SLove}: 
\begin{equation} \label{eq:Sfluct}
S_{\rm Love}^\prime=\frac{i}{2}
\sum_\ell \frac{1}{\ell!}\int d\tau \nabla_\ell\phi_-
H_\ell(i\partial_\tau)\nabla_\ell\phi_-
\end{equation}
where we expect $H_\ell(\omega)\approx \coth\frac{\omega}{2T}
{\rm Im}\,F_\ell(\omega)$ by the fluctuation-dissipation theorem.
Note that since ${\rm Im}\,F_\ell(\omega)\propto \omega$, $H_\ell(\omega)$ is analytic at low frequencies and thus local in time in agreement with EFT logic. (Fluctuation-dissipation may not apply exactly to a black hole formed by gravitational collapse, which is not in equilibrium with its environment and is typically modeled by the so-called Unruh vacuum. We expect a small difference in the near zone at low frequencies, as can be seen using eq.~(5) of \cite{Candelas:1980zt}, where infalling modes are small:, $\overleftarrow{R}_\ell(\omega)\propto (GM\omega)^{2\ell+1}$.)

The action \eqref{eq:Sfluct} does not contribute to the retarded amplitude \eqref{eq:BH exp val} since it cannot connect to external $\phi_-\phi_+$ using the propagators \eqref{eq:props}.
However, it contributes to the  
time-ordered amplitude $\langle \mathcal{T}\, \phi \phi\rangle_\text{BH}=\langle \phi_1\phi_1\rangle_\text{BH}$ since
$\phi_1=\phi_++\frac12 \phi_-$.
This leads to additional diagrams 
\be
S^{\mathcal{T}}-S^\text{ret}\,\supset 
\raisebox{-1.7em}{\includegraphics[width=1.65cm]{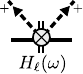}} \,\,.
\ee
These contributions
are enhanced at low frequencies by the Bose-Einstein factor $\sim\frac{2T}{\omega}$ below \eqref{eq:Sfluct}:
\begin{align} \label{eq:F Bose}
   \!\!\! H_{\ell=0}(\omega)\!\approx\!
    \frac{2T}{\omega} {\rm Im }\,F_{\ell=0}(\omega)=4 GM+\mathcal{O}(G^3\omega^2)
\end{align}
where we used the Hawking temperature $T=\frac{1}{8\pi GM}$ and the first nonvanishing Love number from Table \ref{tab:Love}. We stress that while the Hawking temperature represents a small, quantum, energy ($T\propto \hbar$), it is large compared with the energy of the considered wave: $T/\omega\gg 1$.
This is why the worldline action \eqref{eq:Sfluct} starts at order $G$, even though \eqref{eq:SLove} starts at order $G^2$.
It predicts field fluctuations in the near-zone region $GM\ll r\ll \omega^{-1}$ which are approximately frequency-independent and thus appear like white noise on time scales $t\gg r$:
\begin{align}
    \langle \mathcal{T}\phi(r,t)\phi(r',t')\rangle_\text{BH} \approx \frac{H_{\ell=0}(0)}{(4\pi)^2 rr'}\delta(t-t')\,.
\end{align}
The ``fluctuation Love numbers" $H_\ell$  could be measured in principle from decoherence effects which they quantitatively control \cite{Danielson:2022sga,Gralla:2023oya,Wilson-Gerow:2024ljx,Biggs:2024dgp}.

In summary, the linear response of the black hole is captured by a retarded amplitude which admits a simple Born series, a smooth classical limit, and is completely insensitive to Hawking radiation.  In contrast, the low frequency limit of the time-ordered amplitude is dominated by Bose-enhanced fluctuations that could be detected in interference experiments.

\section{Iterated exponential integrals}\label{app:intexample}

Here we exemplify our method to evaluate  the interated integrals \eqref{eq:Irecursion} in the limit $z_{n+1}\to 0$ in terms of multi-zeta values. A distinct method was discussed in \cite{Aminov:2024aan}.
Since the limit in general contains logarithmic divergences the expansion is best organized by introducing the function
\begin{align} \label{eq:Icomplete}
    {\cal I}^{(\eta)}_{j_1\ldots j_n}&\equiv \eta\int_0^\infty \frac{dz_{n+1}}{z_{n_1}}
    (z_{n+1})^\eta\, {\cal I}_{j_1\ldots j_n}(z_{n+1}) 
\end{align}
whose Laurent series up to order $\eta^0$ is in one-to-one correspondence with powers of $\log z_{n+1}$ in the limit.
This is precisely the function given in \eqref{eq:Ieta}. 
For the one-index case the integrals \eqref{eq:Irecursion} give incomplete $\Gamma$ integrals
\begin{align} \label{eq:Isingleindex}
    {\cal I}_{j}(z_2)&= -\int_{z_2}^{\infty} \frac{dz_1}{z_1} e^{2iz_1}  \frac{\log(\frac{z_2}{z_1})^{j-1}}{(j-1)!}
\end{align}
while the complete integral \eqref{eq:Icomplete} evaluates to
\begin{align}
{\cal I}^{(\eta)}_{j}&=
\int_0^\infty \frac{dz_2}{z_2}(z_2)^\eta{\cal I}_{j}(z_2)=
\frac{\Gamma(1{+}\eta)}{(-\eta)^{j}(-2i)^\eta}\,,
\end{align}
in agreement with \eqref{eq:Ieta}. Here the integral can be performed directly by swapping the order and integrating over $z_2$ before $z_1$.
By multiplying \eqref{eq:Isingleindex} by $z^{-\eta}$ and keeping terms of order $\eta^0$ as $\eta\to 0$, one immediately deduces the limit of ${\cal I}_j(z)$, for example:
\begin{align} \label{eq:Isingleindex limit}
    {\cal I}_{1}(z) &\to \log\left(-2i z e^{\gamma_\text{E}}\right) +O(z)\equiv L(z)+{\cal O}(z),\\
    {\cal I}_{2}(z) &\to \frac{L(z)^2}{2}+ \frac{\pi^2}{12} +{\cal O}(z),\\
    {\cal I}_{3}(z) &\to \frac{L(z)^3}{3!}+ \frac{\pi^2}{12}L(z)+\frac{\zeta_3}{3} +{\cal O}(z)\,.
\end{align}
After these leading terms are obtained, subleading powers as $z\to 0$ can be obtained more easily by expanding the integrand in \eqref{eq:Irecursion} and using the simple identity $\int_{z}^\infty=\int_{\delta}^\infty-\int_{\delta}^z$ with $\delta{\to}0$.  All subleading terms here have simple rational coefficients.

In the two-index case the integral \eqref{eq:Irecursion} reads
\begin{equation}\label{eq:Itwoindex}
\!\!\!{\cal I}_{j_1,j_2}(z)\! =\! -\!\!\int_z^\infty \frac{\mathrm{d}z_2}{z_2} e^{-2iz_2} \frac{\log(\frac{z}{z_2})^{j_2{-}1}}{(j_2{-}1)!}
{\cal I}_{j_1}(z_2).
\end{equation}
Notice the alternating sign in the exponent compared with \eqref{eq:Isingleindex}, which is natural from the perspective of how the integrals appear from the Born series.
The regulated complete integral \eqref{eq:Ieta} then reads
\begin{equation}
    {\cal I}_{j_1,j_2}^{(\eta)}\!\! =\! \frac{-\Gamma(1+\eta)}{(-\eta)^{j_2}(-2i)^{\eta}}\!
    \int_0^1\!\! \frac{\mathrm{d}u_1}{1{-}u_1} \frac{\log^{j_1-1}(1{-}u_1)}{(j_1-1)!}\! \left[\frac{u_1}{1{-}u_1}\right]^\eta\nonumber
\end{equation}
which now does not admit a simple closed form.
Nonetheless, the $\eta\to 0$ limit can be computed using HPL technology and we find for example
\begin{align}
    \!\!\!{\cal I}_{1,1}^{(\eta)}&=\frac{1}{(-2i e^{\gamma_\text{E}})^{\eta}}
    \left[\frac{1}{\eta^2}+\frac{\pi^2}{4}+{\cal O}(\eta)\right]\!,
\\
    \!\!\!{\cal I}_{1,2}^{(\eta)}&=\frac{1}{(-2i e^{\gamma_\text{E}})^{\eta}}\left[
\frac{-1}{\eta^3}-\frac{\pi^2}{4\eta}+\frac{\zeta_3}{3}+{\cal O}(\eta)\right]\!,
\\
   \!\!\! {\cal I}_{2,1}^{(\eta)}&=\frac{1}{(-2i e^{\gamma_\text{E}})^{\eta}}\left[
\frac{-1}{\eta^3}-\frac{\pi^2}{12\eta}-\frac{2\zeta_3}{3}+{\cal O}(\eta)\right]\!.
\end{align}
These immediately imply corresponding limits for the incomplete integrals,  for example
\begin{align}
    {\cal I}_{2,1}(z) &\to \frac{L(z)^3}{3!}+ \frac{\pi^2}{12}L(z)-\frac{2\zeta_3}{3} +{\cal O}(z)
\end{align}
with $L(z)$ as in \eqref{eq:Isingleindex limit}.
Again, subleading powers of $z$ are easier to calculate and can be obtained by expanding the integrand in \eqref{eq:Itwoindex},
or alternatively, using the general differential equation
\begin{align}
    \frac{d}{dz}{\cal I}_{\ldots ,j_n}(z)=
\frac{1{-}\delta_{j_n,1}}{z}{\cal I}_{\ldots,j_n-1}(z)
+  \delta_{j_n,1}\frac{e^{\sigma_n z}}{z}{\cal I}_{\ldots}(z)\,.
\end{align}
We conclude this appendix by recording the remaining complete integral of transcendental weight three:
\begin{align}
      {\cal I}_{1,1,1}^{(\eta)}&=\frac{1}{(-2i e^{\gamma_\text{E}})^{\eta}}\left[
\frac{-1}{\eta^3}-\frac{\pi^2}{4\eta}+\frac{7\zeta_3}{3}+{\cal O}(\eta)\right].
\end{align}
\onecolumngrid
\section{Results up to $O(G^7)$}\label{app:results}
In this Appendix we collect lengthy results. We give the dimensionally regulated phase-shifts obtained from the iterative procedure described in the main text up to $\mathcal{O}(G^7)$, including for simplicity only long-distance gravitational contributions, and then give data related to the renormalization procedure and renormalized Love numbers. 

The IR divergences are treated by introducing a hard cutoff $R$ in the iterated Born integrals (\ref{eq:psiBorn}), and it can be easily shown that dimensional regularization is recovered by the replacement
\begin{equation}
\label{eq:harddimreg}
\log R \mapsto   - {1 \over 2 \epsilon_{IR}} +  \log \bar \mu_{\text{IR}}^{-1}\,,
\end{equation}
with $\bar \mu_\text{IR}^2 = \mu_\text{IR}^2 4\pi e^{\gamma_{E}-1}$.
The long distance contribution to the phase shift is given by
\begin{align}\nonumber
\delta_\mathrm{Grav}^{\ell=0}\,(\omega)&=GM\omega\left(\frac{1}{\epsilon}+\log{\frac{4\omega^2}{\bar\mu_{\mathrm{IR}^2}}} -1+2\gamma_E\right)+\frac{11\pi}{3}\left(GM\omega\right)^2\\\nonumber&+\left(GM\omega\right)^3\left(\frac{2}{\epsilon }-6 \log \left(\frac{4 \omega ^2}{\bar\mu^2}\right)+\frac{62}{3}+\frac{22 \pi ^2}{9}-\frac{8 }{3}\zeta_3 \right)\\\nonumber
&+\left(GM\omega\right)^4\left(\frac{4 \pi }{\epsilon }-16 \pi  \log \left(\frac{4 \omega ^2}{\bar\mu ^2}\right)+\frac{6799 \pi }{135}\right)+\left(GM\omega\right)^5\Bigg[\frac{11}{3 \epsilon ^2}+\frac{1}{3\epsilon}\left(194+8 \pi ^2-55 \log \left(\frac{4 \omega ^2}{\bar\mu ^2}\right)\right)\\\nonumber
&\;\;\;\;\;\;+\frac{275}{6} \log ^2\left(\frac{4 \omega ^2}{\bar\mu^2}\right)-\frac{10}{3} \left(97+4 \pi ^2\right) \log \left(\frac{4 \omega ^2}{\bar\mu ^2}\right)+\frac{105458}{135}+\frac{3407 \pi ^2}{1620}-\frac{616 }{9}\zeta_3-\frac{88 \pi ^4}{135}+\frac{32}{5} \zeta_5\Bigg]\\\nonumber
&+(GM\omega)^6\Bigg[\frac{22 \pi }{3 \epsilon ^2}-\frac{4 \pi}{3\epsilon}  \left(33 \log \left(\frac{4 \omega ^2}{\bar\mu^2}\right)-82\right)+132 \pi  \log ^2\left(\frac{4 \omega ^2}{\bar\mu ^2}\right)-656 \pi  \log \left(\frac{4 \omega ^2}{\bar\mu^2}\right)\\\nonumber
&\;\;\;\;\;\;-\frac{176}{3}  \pi^3-192 \pi  \zeta_3-33 \pi ^3+\frac{52073626 \pi }{42525}\Bigg]\\\nonumber
&+(GM\omega)^7\Bigg[ \frac{157}{27 \epsilon ^3} +\frac{1}{{405 \epsilon ^2}}\left(-16485 \log \left(\frac{4 \omega ^2}{\bar \mu ^2}\right)+1980 \pi ^2+56663\right)\\\nonumber
&\;\;\;\;\;\;+\frac{1}{{8100 \epsilon }}\Bigg(1153950 \log ^2\left(\frac{4 \omega ^2}{\bar\mu ^2}\right)-140 \left(56663+1980 \pi ^2\right) \log \left(\frac{4 \omega ^2}{\bar\mu^2}\right)-2138400 \zeta_3\\\nonumber
&\;\;\;\;\;\;-5760 \pi ^4-314325 \pi ^2+17790348\Bigg)\\\nonumber
&\;\;\;\;\;\;-\frac{53851}{162} \log ^3\left(\frac{4 \omega ^2}{\bar\mu^2}\right)+\frac{49}{810} \left(56663+1980 \pi ^2\right) \log ^2\left(\frac{4 \omega ^2}{\bar\mu ^2}\right)\\\nonumber
&\;\;\;\;\;\;+\frac{7 \log \left(\frac{4 \omega ^2}{\bar\mu^2}\right)}{2700}\left(712800 \zeta_3+1920 \pi ^4+104775 \pi ^2-5930116\right) -\frac{320}{3} \pi ^2 \zeta_3+\frac{2048 \zeta_5}{9}-\frac{128 \zeta_7}{7}\\\nonumber
&\;\;\;\;\;\;-\frac{266929 \zeta_3}{45}+\frac{18572969659}{637875}+\frac{704 \pi ^6}{2835}-\frac{78613 \pi ^4}{675}-\frac{937910047 \pi ^2}{510300}\Bigg]\,,
\end{align}

\begin{align}\nonumber
\delta_\mathrm{Grav}^{\ell=1}\,(\omega)&=GM\omega\left(\frac{1}{\epsilon}+\log{\frac{4\omega^2}{\bar\mu_{\mathrm{IR}^2}}} -3+2\gamma_E\right)+(GM\omega)^2\frac{19 \pi }{15}\\\nonumber
&+(GM\omega)^3\left(\frac{38 \pi ^2}{45}-\frac{8 \zeta_3}{3}\right)+(GM\omega)^4\frac{78037 \pi }{23625}\\\nonumber
&+(GM\omega)^5\left[\frac{1}{9 \epsilon }-\frac{5}{9} \log \left(\frac{4 \omega ^2}{\bar\mu^2}\right)+\frac{3022}{675}+\frac{156074 \pi ^2}{70875}-\frac{152 \pi ^4}{675} +\frac{2888 \zeta_3}{225}+\frac{32 \zeta_5}{5}\right]\\\nonumber
&+(GM\omega)^6\left[ \frac{2 \pi }{9 \epsilon }-\frac{4}{3} \pi  \log \left(\frac{4 \omega ^2}{\bar\mu^2}\right)+\frac{909340802 \pi }{37209375}\right]\\\nonumber
&+(GM\omega)^7\Bigg[\frac{19}{405 \epsilon ^2}+\frac{1}{6075 \epsilon }\left( -1995 \log \left(\frac{4 \omega ^2}{\bar\mu ^2}\right)+900 \pi ^2+16739\right)+\frac{931}{810} \log ^2\left(\frac{4 \omega ^2}{\bar\mu^2}\right)\\\nonumber
&-\frac{7 \left(16739+900 \pi ^2\right) \log \left(\frac{4 \omega ^2}{\bar\mu ^2}\right)}{6075}+\frac{93540397}{1063125}+\frac{7014260791 \pi ^2}{446512500}-\frac{8896 \pi ^4}{39375}\\\nonumber
&+\frac{1216 \pi ^6}{14175}+\frac{18368248 \zeta_3}{354375}-\frac{23104 \zeta_5}{225}-\frac{128 \zeta_7}{7}\Bigg]\,,
\end{align}

\begin{align}\nonumber
    \delta_\mathrm{Grav}^{\ell=2}\,(\omega)&=GM\omega\left(\frac{1}{\epsilon}+\log{\frac{4\omega^2}{\bar\mu_{\mathrm{IR}^2}}} -4+2\gamma_E\right)+(GM\omega)^2\frac{79 \pi }{105}+(GM\omega)^3\left( \frac{158 \pi ^2}{315}-\frac{8 \zeta_3}{3}\right)\\\nonumber
    &+(GM\omega)^4\frac{708247 \pi }{1157625}
    +(GM\omega)^5\left[-\frac{16903}{33075}+\frac{1416494 \pi ^2}{3472875}-\frac{632 \pi ^4}{4725}+\frac{49928 \zeta_3}{11025}+\frac{32 \zeta_5}{5}\right]\\\nonumber
    &+(GM\omega)^6\frac{173197070512 \pi }{140390971875}+(GM\omega)^7\Bigg[ \frac{8}{6075 \epsilon }-\frac{56 \log \left(\frac{4 \omega ^2}{\bar\mu ^2}\right)}{6075}-\frac{649093993}{2187911250}\\\nonumber
    &+\frac{346394141024 \pi ^2}{421172915625}-\frac{3904 \pi ^4}{118125}+\frac{5056 \pi ^6}{99225}+\frac{895224208 \zeta_3}{121550625}-\frac{399424 \zeta_5}{11025}-\frac{128 \zeta_7}{7}\Bigg]\,.
    \end{align}
These expressions constitute a prediction for six-loop gravitational Feynman diagrams integrated over Legendre polynomials.
The full phase shift is obtained from the above plus the short-distance effects controlled by the Love numbers $F_\ell$, which to leading order enter as 
\begin{equation}
S^{\ell}_\text{Full}(\omega) = e^{2i\delta^\ell_{\rm Grav}(\omega)} + i
F_\ell(\omega)
\frac{
 \mu^{2\epsilon} \omega^{2\ell+1-2\epsilon}
 2^{-\ell-2 + 2\epsilon}}{\pi^{\frac12-\epsilon}\Gamma(\frac32{-}\epsilon{+}\ell)}
+ \mathcal{O}(F_\ell^2)\,.
\end{equation}
The $1/\epsilon$ poles in $\delta_\text{Grav}$ must cancel against poles from the Love number contribution. As explained in the main text, the cancellation becomes transparent once the boundary condition \eqref{eq:psi0near} is written in
terms of renormalised coefficients $(B_{\rm reg},B_{\rm irr})^T=Z(\bar{B}_{\rm reg},\bar{B}_{\rm irr})^T$,
such that the wave solution expressed in terms of $\bar{B}_\text{reg}$ and $\bar{B}_\text{irr}$ is finite
as $d\to4$.
We use a minimal subtraction scheme in which the $Z$ factor contains powers of $\bar{G}=GMn_d$ times pure inverse powers of $\epsilon$, for example for $\ell=0$ we obtain
\begin{align}
Z^{\ell=0} = \begin{pmatrix}
    1+\frac{11(\bar{G}\omega)^2}{6\epsilon} \quad&\quad
\frac{\bar{G}\omega^2}{\epsilon} + \frac{14(\bar{G}\omega)^3\omega}{3\epsilon}
\\[0.2cm]
-\frac{2\bar{G}^3\omega^2}{\epsilon}
& 1 - \frac{11\bar{G}^2\omega^2}{6\epsilon}
\end{pmatrix}+\mathcal{O}(G^4)\qquad
(\bar{G}\equiv GMn_d)\,.
\end{align}
In terms of renormalized quantities, the boundary condition \eqref{eq:psi0near} in the perturbative near zone ($GM\ll r \ll \omega^{-1}$) depends logarithmically on $\mu$, for example 
\begin{align}\label{psi0d4}
    \lim_{d\to 4} \frac{1}{r^{\frac{d-2}{2}}}\psi^{\ell=0}(r) &= \bar{B}_{\text{reg}}\left[
    1 +(\bar{G}\omega)^2 \left(\frac{16}{9} {-} \frac{22}{3} \log(r\mu)\right) + 
\frac{\bar{G}^3\omega^2}{ r}\left(\frac{68}{3}{+}8\log(r\mu)\right) \right]
\\&   + \bar{B}_\text{irr}
\left[\frac{1}{r} + \frac{(\bar{G}\omega)^2}{r}\left(19{+} \frac{22}{3}\log(r\mu)\right)
-\bar{G}\omega^2\left(\frac{1}{2} {+} 4\log(r\mu)\right)
-\bar{G}^3\omega^4
\left(\frac{1205}{108} {+} \frac{433}{9} \log(r\mu)\right)\right] \nonumber\\ &+ \mathcal{O}(\bar{G}^4)+\mbox{other powers}\nonumber
\end{align}
where we omitted terms with powers of $r$ distinct from $\ell$ and $-\ell-1$ (the wavefunction in the considered region is a double expansion in $\frac{\bar{G}}{r}$ and $\omega r$). By matching with the GR solution in this region it enables the extraction of precise but $\mu$ dependent values of $\bar{B}$'s and thus Love numbers in a specific minimal subtraction scheme via \eqref{eq:BLove}.
The logarithms could be predicted using the wave equation in $d=4$ but the constant terms are specific to the particular minimal subtraction scheme picked by the worldline EFT.
As noted in the text, $\bar{B}$ depends on $\mu$ through the anomalous dimension matrix
\begin{equation}\label{eq:defgamma}
    \gamma^\ell = (Z^\ell)^{-1}\mu\frac{d}{d\mu}Z^\ell + (Z^\ell)^{-1}\begin{pmatrix} 
    0&0\\0&2\epsilon
    \end{pmatrix}Z^\ell\,.
\end{equation}
Here $\mu\frac{d}{d\mu}=\mu\frac{\partial}{\partial \mu}+\beta(G)\frac{\partial}{\partial G}$ is the Callan-Symanzik operator where
$\beta(G)\equiv \mu\frac{d}{d\mu}G=-2\epsilon G$ (ensuring that the metric \eqref{eq:metric} is invariant). Similarly the second term in \eqref{eq:defgamma} comes from the $\mu^{2\epsilon}$ in \eqref{eq:SLove} and \eqref{eq:psi0near}.
In minimal subtraction the anomalous dimension is exactly linear in $\epsilon$, and we find explicitly up to $\mathcal{O}(G^8)$ error:
\begin{align}\nonumber
    \gamma^{(0)}=\begin{pmatrix}
   -\left[\frac{22 (\bar G \omega)^2}{3}+\frac{7118 (\bar G \omega)^4}{135}+\frac{22329152 (\bar G \omega)^6}{42525}\right] & \,\,\,\,\,-\omega\left[ 4\bar G\omega+\frac{112 (\bar G \omega)^3}{3}+\frac{125402 (\bar G \omega)^5}{405}+\frac{6707817458(\bar G \omega)^7}{1913625}\right]\\[0.4cm]
  \bar G\left[8 (\bar G \omega)^2+72 (\bar G \omega)^4+ \frac{520408 (\bar G \omega)^6}{675}\right]& \frac{22(\bar G \omega)^2}{3}+\frac{7118 (\bar G \omega)^4}{135}+\frac{22329152 (\bar G \omega)^6}{42525}+2 \epsilon
    \end{pmatrix}\,,
\end{align}
\begin{align}\nonumber
    \gamma^{(1)}=\begin{pmatrix}
 {-}\!\left[ \frac{38 (\bar G \omega)^2}{15}{+}\frac{156074 (\bar G \omega)^4}{23625} {+} \frac{1504469104 (\bar G \omega)^6}{37209375}\right]\!\!\!\!\!\!& \,\,\,\,\,-\omega^3\!\left[\frac{4 \bar G \omega}{9}{+}\frac{7648 (\bar G \omega)^3}{2025}{+}\frac{463800938(\bar G \omega)^5}{22325625}{+} \frac{4484979447502 (\bar G \omega)^7}{35162859375}\right]\\[0.4cm]
  \bar{G}^3\left[8(\bar G \omega)^2{+}\frac{5804 (\bar G \omega)^4}{75}\right]& +\frac{38 (\bar G \omega)^2}{15}{+}\frac{156074 (\bar G \omega)^4}{23625}{+}\frac{1504469104 (\bar G \omega)^6}{37209375}+2 \epsilon
    \end{pmatrix}\,,
\end{align}

\begin{align}\nonumber
    \!\!\!\!\!\!\!\!\!\gamma^{(2)}=\begin{pmatrix}
{-}\!\left[\frac{158(\bar G \omega)^2}{105}{+}\frac{1416494(\bar G \omega)^4}{1157625}{+}  \frac{346394141024 (\bar G \omega)^6}{140390971875}\right]\!\!\!\!\!\!\!& \,\,\,\,\,-\omega^5\!\!\left[\frac{4(\bar G \omega)}{225}{+}\frac{343552 (\bar G \omega)^3}{2480625}{+}\frac{14735507852(\bar G \omega)^5}{27348890625}{+}\frac{12876600793881974 (\bar G \omega)^7}{7296820763203125}\right]\\[0.4cm]
  \frac{32\bar G^7\omega^2}{9}& \frac{158 (\bar G \omega)^2}{105}{+}\frac{1416494(\bar G \omega)^4}{1157625}{+}\frac{346394141024 (\bar G \omega)^6}{140390971875}+2 \epsilon
    \end{pmatrix}\,.
\end{align}
All coefficients are rational numbers since the $Z$ matrices can be obtained directly from the near-zone solution. From these, the RG equations \eqref{eq:RGE} are directly read off. For example in $d=4$ we have explicitly
\begin{align}\nonumber
\qquad \mu \frac{d}{d\mu}\bar F_0(\omega)& = -4 \pi GM\left[8 (GM \omega)^2+72 (G M\omega)^4+ \frac{520408 (GM\omega)^6}{675}\right]\\
& -2\bar F_0(\omega)\left[\frac{22 (GM \omega)^2}{3}+\frac{7118 (GM \omega)^4}{135}+\frac{22329152 (GM\omega)^6}{42525}\right]\\\nonumber
&- \frac{\bar F_0(\omega)^2}{4\pi}
\omega\left[ 4GM\omega+\frac{112 (G M\omega)^3}{3}+\frac{125402 (GM \omega)^5}{405}+\frac{6707817458(GM \omega)^7}{1913625}\right].
\end{align}
\begin{align}\nonumber
 \mu\frac{d}{d\mu}\bar F_1(\omega)& =-\frac{4\pi }{3}G^3 M^3\left[8 (G M \omega )^2+\frac{5804}{75} (G M \omega )^4\right]\\
& -2\bar F_1(\omega)\left[\frac{38}{15} (GM\omega)^2 +\frac{156074 }{23625}(GM\omega)^4+\frac{1504469104  }{37209375}(GM\omega)^6 \right]\\\nonumber
&-\frac{3\bar F_1(\omega)^2}{4\pi}\omega^3\left[\frac{4  }{9}G M \omega+\frac{7648 }{2025}(GM\omega)^3+\frac{463800938 }{22325625}(GM\omega)^5+\frac{4484979447502 }{35162859375}(GM\omega)^7\right]
\end{align}
\begin{equation}
\begin{aligned}
     \mu \frac{d}{d\mu}\bar F_2(\omega)& =-\frac{4\pi}{15}G^5 M^5\left[\frac{32}{9} (G M \omega) ^2\right]\\
& -2\bar F_2(\omega)\left[\frac{158}{105} (G M \omega) ^2+\frac{1416494 }{1157625}(GM\omega)^4+\frac{346394141024 }{140390971875}(GM\omega)^6\right]\\
&-\frac{15\bar F_2(\omega)^2}{4\pi}\omega^5\Bigg[\frac{4 }{225}G M \omega +\frac{343552 }{2480625}(GM\omega)^3+\frac{14735507852}{27348890625}(GM\omega)^5\\
&\qquad+\frac{12876600793881974 }{7296820763203125}(GM\omega)^7\Bigg]
\end{aligned}\end{equation}
The running of individual Love numbers $\bar{C}_{\ell,n}(\mu)$ is determined by Taylor-expanding the expression above in terms of $\omega$, using the relation \eqref{eq:Fl}. We find that the static Love numbers, which are leading in $\omega$, do not evolve. In contrast, the dynamical Love numbers, which are sub-leading in $\omega$, exhibit running and become non-linearly coupled between each other due to the quadratic nature of the RGE.

Finally, we list the renormalised Love numbers up to $\mathcal{O}(G^7)$, while the results up to $\mathcal{O}(G^5)$ for the bare coefficients are collected in Table \ref{tab:Love}.
\begin{align}
   \frac{1}{4\pi} \bar F_0(\omega) &= i R_s^2 \omega - R_s^3 \omega^2\left( \log (\mu R_s)+\frac{11}{6}\right)- i R_s^4 \omega ^3\left(\frac{11}{3} \log (\mu R_s)+\frac{10}{9}-2 \zeta_2 \right)\nonumber\\
   & + R_s^5 \omega^4\left (\frac{233}{36} \log (\mu  R_s)+\frac{11}{6} \log ^2(\mu  R_s) -\frac{559}{1080}-\frac{22 \zeta_2}{3}-2 \zeta_3\right)\nonumber\\
   & +i R_s^6\omega^5 \left[\left(\frac{289}{60}-\frac{22 \zeta_2}{3}\right) \log (\mu  R_s)+ \frac{157}{18} \log ^2(\mu  R_s)+\frac{469}{1200}-\frac{233 \zeta_2}{18}-\frac{4 \zeta_2^2}{5}-\frac{22 \zeta_3}{3}\right]\\
   &-R_s^7 \omega^6\Bigg[-\left(\frac{314 \zeta_2}{9}+\frac{22 \zeta_3}{3}+\frac{47227}{32400}\right) \log (\mu R_s)+ \frac{2642}{135} \log ^2(\mu R_s)+\frac{157}{54} \log ^3(\mu R_s)\nonumber\\
   & \qquad\qquad +\frac{23365109}{10206000} -\frac{289 \zeta_2}{30}+\frac{22 \zeta_2^2}{5}-\frac{92 \zeta_3}{9}-2 \zeta_5\Bigg]\nonumber
\end{align}
\begin{equation}
\begin{aligned}
    \frac{3}{4\pi}\bar F_1(\omega) &= \frac{1}{4} i R_s^4 \omega - R_s^5 \omega^2\left( \frac{1}{4} \log (\mu  R_s)-\frac{1}{160}\right)-iR_s^6\omega^3\left(\frac{19}{60} \log (\mu  R_s)-\frac{1283}{1800}-\frac{\zeta_2}{2} \right)\\
    &+ R_s^7 \omega^4\left[-\frac{49}{80} \log (\mu  R_s)+\frac{19}{120} \log ^2(\mu R_s)+\frac{6247}{30240}-\frac{19 \zeta_2}{30} -\frac{\zeta_3}{2}\right]
\end{aligned}\end{equation}
\begin{equation}
    \frac{15}{4\pi}\bar F_2(\omega) = \frac{1}{36} i R_s^6- R_s^7 \omega^2 \left(\frac{1}{36}  \log (\mu  R_s)-\frac{83 }{5670}\right) 
\end{equation}

\vspace{-0.9cm}
\twocolumngrid
\bibliography{Born}

\end{document}